%% file: main.tex
\documentclass{moriond}

\def\be{\begin{equation}}
\def\ee{\end{equation}}
\def\bea{\begin{eqnarray}}
\def\eea{\end{eqnarray}}

\newcommand\sgr{SGR~1935+2154}
\input{macros}

\newcommand{\Photo}{\includegraphics[height=35mm]{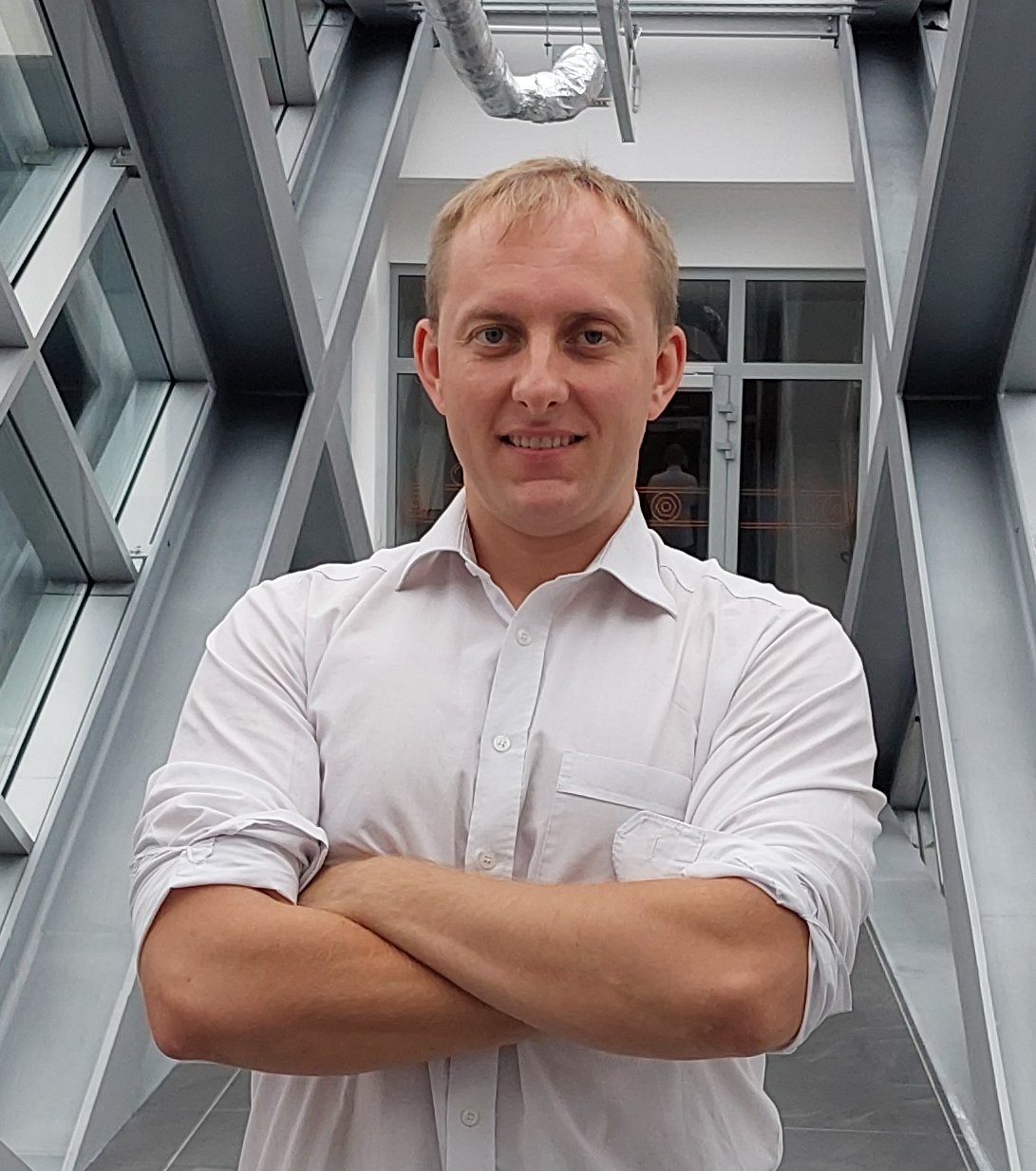}}

\begin{document}
\vspace*{4cm}
\title{Targeted searches for gravitational waves from SN 2023ixf and SGR 1935+2154}

\author{Marek J. Szczepa\'{n}czyk, marek.szczepanczyk@fuw.edu.pl\\
\url{https://www.fuw.edu.pl/~mszczepanczyk/}\\
on behalf of the LIGO, Virgo and KAGRA Scientific Collaborations}
\address{Faculty of Physics, University of Warsaw, Ludwika Pasteura 5, 02-093 Warszawa, Poland.}

\maketitle\abstracts{
The fourth observing run of Advanced LIGO, Advanced Virgo, and KAGRA has provided so far over 200 new gravitational-wave candidates, and it is still ongoing. A few results from this run are published and in this proceeding, we summarize the latest targeted search for GWs from SN~2023ixf and consider predictions for future searches. We also summarize the targeted search for GWs from a fast radio burst source SGR 1935+2154.
}

\section{Introduction}

The discovery of gravitational waves (GWs) from binary black holes~\cite{Abbott:2016blz} opened a new window to explore the Universe. Since then, the Advanced LIGO~\cite{TheLIGOScientific:2014jea} and Advanced Virgo~\cite{TheVirgo:2014hva} have observed nearly 100 detections of compact binary mergers (neutron stars and/or black holes)~\cite{GWTC1,GWTC2,GWTC2.1,GWTC3} during the first three observing runs (O1-O3) and over 200 candidates during the fourth observing run (O4)\footnote{The list of O4 GW candidates is available at: \url{https://gracedb.ligo.org/superevents/public/O4/}.}. A GW burst source has not yet been observed and 10 years after GW150914~\cite{Abbott:2016blz} an interest is growing into observing a non-binary source of GWs by LIGO, Virgo, and KAGRA~\cite{Aasi:2013wya} (LVK) collaboration. Core-collapse supernovae (CCSNe) are one of the most promising candidates\footnote{See the Symposium on CCSNe (July 21-25, 2025): \url{https://indico2.fuw.edu.pl/event/17/overview}.}. In this proceeding, we briefly summarize the latest targeted search for GWs from SN~2023ixf~\cite{SN2023ixf} and provide predictions for future searches. We also summarize the targeted search for GWs from a fast radio burst (FRB) source SGR~1935+2154~\cite{FRB}.

\section{Gravitational-wave search from SN 2023ixf}

CCSNe mark the violent deaths of massive stars (above 8 Sun's masses) culminating in the formation of neutron stars or black holes. These events are multimessenger astrophysical phenomena, historically observed across the electromagnetic spectrum and, in the case of SN 1987A, also through low-energy neutrinos~\cite{hirata:87,bionta:87,1987ESOC...26..237A}. GWs from CCSNe have not yet been detected. Despite the decades-long effort, the explosion mechanism(s) are not yet fully understood, see reviews~\cite{Janka:2012wk,Mezzacappa_2024}.

While a Galactic or near-Galactic CCSN is the most promising to discover GWs~\cite{Szczepanczyk:2021bka}, the targeted searches with CCSNe at the Mpc range allow us to systematically constrain the CCSN engine. The methodology was established with the initial detector data~\cite{Abbott:2016tdt}. The first observational constraints of the CCSN engine were achieved with the O1 and O2 data and SN~2017eaw~\cite{Abbott:2019pxc}. Later, the O3 data did not allow better constraints\cite{Szczepanczyk:2023ihe}, but a broader results interpretation was provided.

\begin{figure}
\centerline{\includegraphics[width=0.7\linewidth]{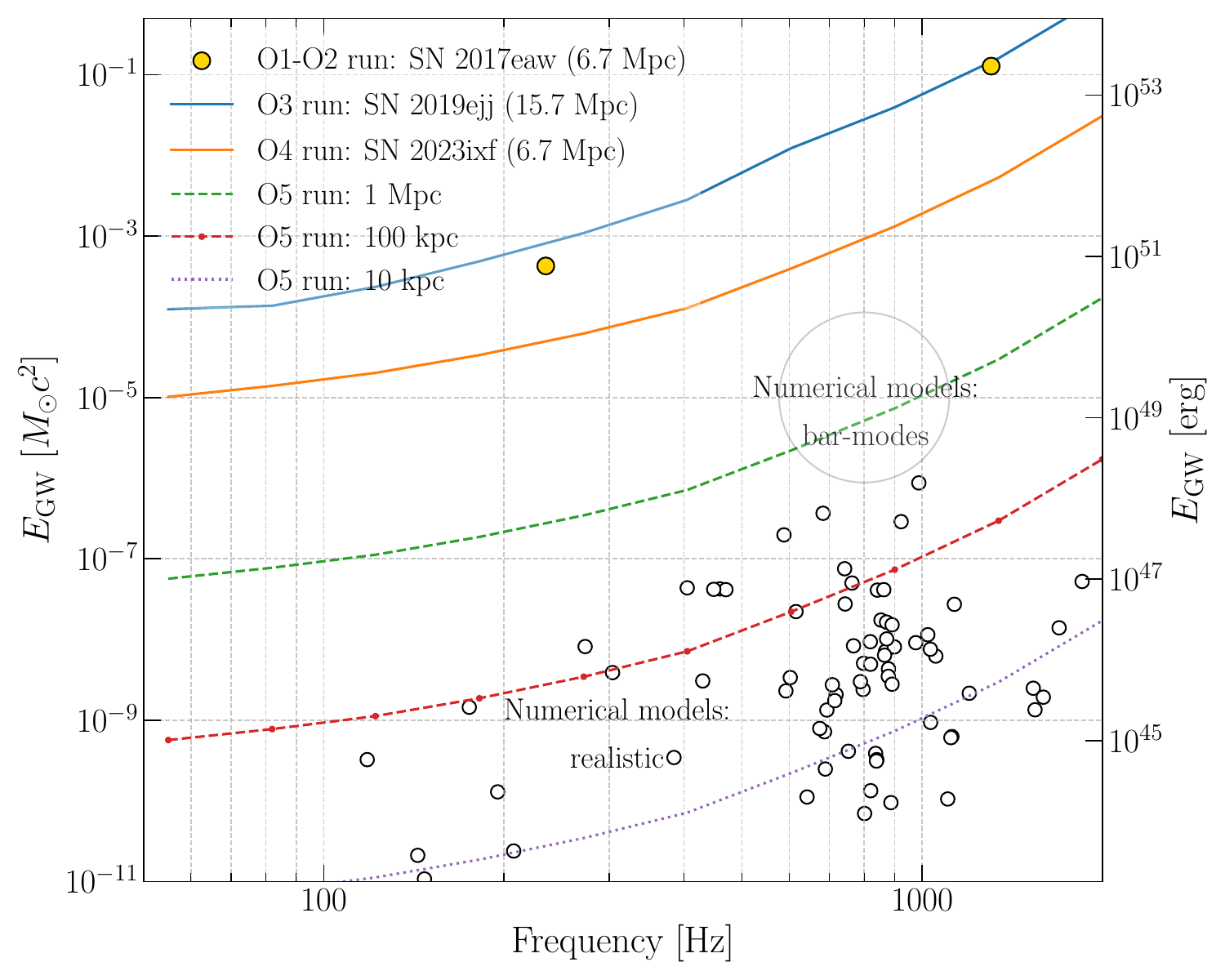}}
\caption{Upper limits on emitted GW energy ($E_\mathrm{GW}$). 
The results with SN~2023ixf are around an order of magnitude more stringent than those obtained with SN~2017eaw. Future nearby CCSN and better detector sensitivity promise more stringent upper limits until a GW discovery that has expected GW energies much smaller than the current limits (white circles, see~\protect\cite{Szczepanczyk:2021bka} for an overview).}
\label{fig:sn}
\end{figure}

SN 2023ixf was discovered on 2023 May 19 by~\cite{2023TNSTR1158....1I} in M101 galaxy at the distance of 6.7\,Mpc~\cite{SN2023ixf}. The time of the expected GW signal was estimated to be before the O4 started. The search was conducted with coherent WaveBurst~\cite{cwb,Klimenko:2022nji} that did not yield any significant GW candidate. While lacking GWs, the calculated upper limits on GW emission became more stringent. Constraints on GW energy, GW luminosity, and GW proto-neutron star ellipticity have improved. In particular, Figure~\ref{fig:sn} shows an upper limit on the energy emitted in GWs, assuming a narrow-band emission (e.g. bar mode instability). The upper limit on GW energy emission improved by around an order of magnitude with SN~2023ixf compared to the results obtained with SN~2017eaw with O1-O2 data (based on the 50\% detection efficiency). Similarly, the upper limits on GW luminosity also improved up to around an order of magnitude.

These results indicate that we do not expect a GW detection for realistic numerical models with CCSNe at Mpc distances. However, the current GW energy upper limits are around an order of magnitude less stringent than the GW energies from more extreme numerical bad-mode models, for example, low rotational kinetic energy over gravitational potential energy ratio (low-$\mathrm{T/|W|}$)~\cite{Shibagaki:20,Bugli:2019rax}. It is therefore feasible to constrain in O5 the more extreme numerical models. The closer CCSNe, the better upper limits we obtain until a GW discovery.

\section{Gravitational-wave search from SGR~1935+2154}

Fast Radio Bursts (FRBs) are energetic ($10^{37}-10^{42}$ erg) millisecond-scale radio transients. Some of them are repeating, and some of them are not. The first FRB was discovered in 2007~\cite{Lorimer:2007qn}, and the number of observations grew rapidly. While the emission mechanism is unknown, they are most likely associated with neutron stars. So far, all known FRBs have originated from outside of the Milky Way. The notable exceptions are repeating FRBs from SGR~1935+2154 magnetar located 6.6\,kpc away. These are the only known FRBs associated with a source.

\begin{figure}
\centerline{\includegraphics[width=0.7\linewidth]{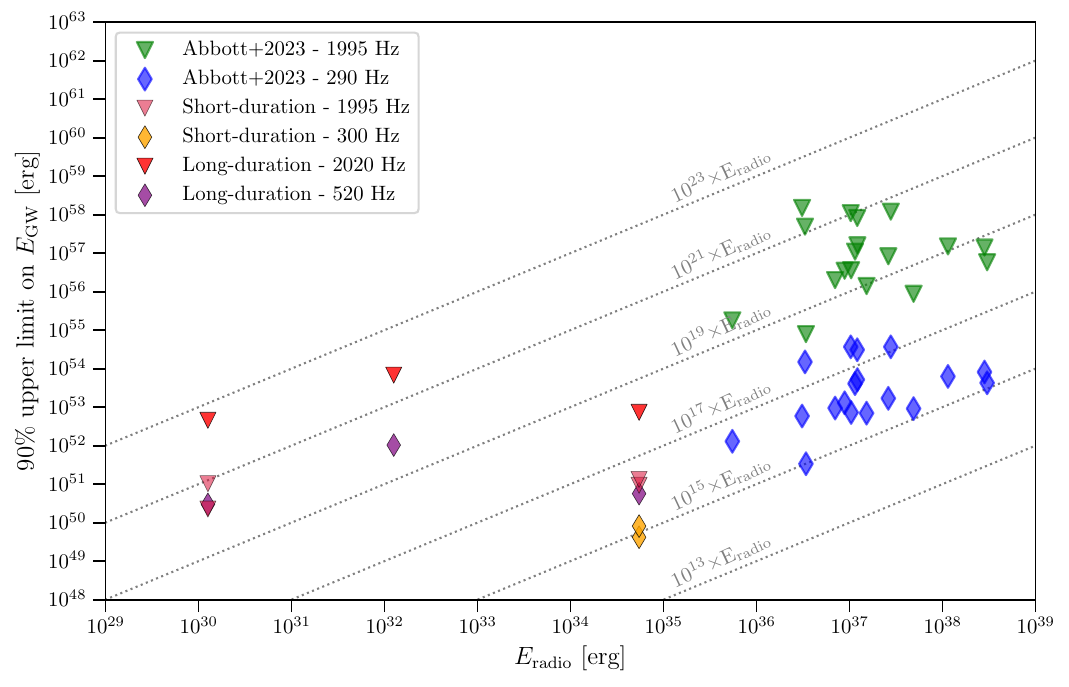}}
\caption{Upper limits on emitted GW energy ($E_\mathrm{GW}$) and energy of radio bursts ($E_\mathrm{radio}$), reproduced from~\protect\cite{FRB}. The upper limits on GW energy have improved 5 orders of magnitude thanks to a close distance of \sgr, but the constraints on the GW to radio energy ratio improved only slightly.}
\label{fig:frb}
\end{figure}

The FRBs from SGR~1935+2154 happened between O3 and O4, and only GEO600~\cite{Luck:2010rt,Affeldt:2014rza,Dooley:2015fpa} data was available at that time. While the overall sensitivity of GEO600 is worse than for LIGO and Virgo, the the sensitivity at 2\,kHz (expected oscillations of neutron stars) is only around an order of magnitude worse making it feasible to perform a search. Therefore, the search was performed for short-duration ($<1$\,s, X-pipeline~\cite{Sutton:2009gi,Was:2012zq}) and long-duration (1-10\,s, PySTAMP~\cite{2021PhRvD.104j2005M}) GW transients. 7 data periods were analyzed (4 FRBs and 3 X-rays). No significant GW candidate was found, but the constraints on GW energy emission improved 5 orders of magnitude with respect to the previous search with O3 data~\cite{LIGOScientific:2022jpr} thanks to a short distance of \sgr. However, the upper limits on ratio between emitted GW energy ($E_\mathrm{GW}$) and energy of radio bursts ($E_\mathrm{radio}$) improved only slightly, see Figure~\ref{fig:frb} (reproduced from~\cite{FRB}).

\section{Conclusions}

The targeted searches for GWs from SN~2023ixf and SGR~1935+2154 are recent results published during O4. The upper limits on emitted GW energy have improved by an order of magnitude for CCSNe and by 5 orders of magnitude for FRB sources. These results are part of the process to systematically constrain the properties of the CCSN engine and FRB emission mechanism, respectively, and prepare for the discovery of a GW burst.

\section*{Acknowledgments}

Acknowledgments may be found in https://dcc.ligo.org/LIGO-P2100218/public. This material is based upon work supported by NSF’s LIGO Laboratory, which is a major facility fully funded by the National Science Foundation. The authors gratefully acknowledge the support and help of the LVK Burst group. M.S. acknowledges Polish National Science Centre Grants No. UMO-2023/49/B/ST9/02777 and No. 2024/03/1/ST9/00005, and the Polish National Agency for Academic Exchange within Polish Returns Programme Grant No. BPN/PPO/2023/1/00019.

\section*{References}
\bibliography{main.bib}


\end{document}

%% file: macros.tex
\renewcommand{\today}{\number\day\space\ifcase\month\or
  January\or February\or March\or April\or May\or June\or
  July\or August\or September\or October\or November\or December\fi
  \space\number\year}

\def\be{\begin{equation}}
\def\ee{\end{equation}}
\def\bi{\begin{itemize}} 
\def\ei{\end{itemize}}
\def\ben{\begin{enumerate}}
\def\een{\end{enumerate}}


%% file: main.bbl
\begin{thebibliography}{10}

\bibitem{Abbott:2016blz}
B.P. Abbott et~al.
\newblock {Observation of Gravitational Waves from a Binary Black Hole Merger}.
\newblock {\em Phys. Rev. Lett.}, 116(6):061102, 2016.

\bibitem{TheLIGOScientific:2014jea}
J.~Aasi et~al.
\newblock {Advanced LIGO}.
\newblock {\em Class. Quant. Grav.}, 32:074001, 2015.

\bibitem{TheVirgo:2014hva}
F.~Acernese et~al.
\newblock {Advanced Virgo: a second-generation interferometric gravitational
  wave detector}.
\newblock {\em Class. Quant. Grav.}, 32(2):024001, 2015.

\bibitem{GWTC1}
B.~P. Abbott et~al.
\newblock {GWTC-1: A Gravitational-Wave Transient Catalog of Compact Binary
  Mergers Observed by LIGO and Virgo during the First and Second Observing
  Runs}.
\newblock {\em Phys. Rev. X}, 9(3):031040, 2019.

\bibitem{GWTC2}
R.~Abbott et~al.
\newblock {GWTC-2: Compact Binary Coalescences Observed by LIGO and Virgo
  During the First Half of the Third Observing Run}.
\newblock {\em Physical Review X}, 11:021053, 2021.

\bibitem{GWTC2.1}
R.~{Abbott}, T.~D. {Abbott}, F.~{Acernese}, K.~{Ackley}, C.~{Adams},
  N.~{Adhikari}, R.~X. {Adhikari}, V.~B. {Adya}, C.~{Affeldt}, D.~{Agarwal},
  and et~al.
\newblock {GWTC-2.1: Deep extended catalog of compact binary coalescences
  observed by LIGO and Virgo during the first half of the third observing run}.
\newblock {\em \prd}, 109(2):022001, January 2024.

\bibitem{GWTC3}
R.~Abbott et~al.
\newblock {GWTC-3: Compact Binary Coalescences Observed by LIGO and Virgo
  during the Second Part of the Third Observing Run}.
\newblock {\em Phys. Rev. X}, 13(4):041039, 2023.

\bibitem{Aasi:2013wya}
B.P. Abbott et~al.
\newblock {Prospects for Observing and Localizing Gravitational-Wave Transients
  with Advanced LIGO, Advanced Virgo and KAGRA}.
\newblock {\em Living Rev. Rel.}, 21(1):3, 2018.

\bibitem{SN2023ixf}
A.~G. Abac et~al.
\newblock {Search for gravitational waves emitted from SN 2023ixf}.
\newblock {\em Astrophys. J.}, 985(2):183, May 2025.

\bibitem{FRB}
A.~G. Abac et~al.
\newblock {A Search Using GEO600 for Gravitational Waves Coincident with Fast
  Radio Bursts from SGR 1935+2154}.
\newblock {\em Astrophys. J.}, 977(2):255, 2024.

\bibitem{hirata:87}
K.~{Hirata}, T.~{Kajita}, M.~{Koshiba}, M.~{Nakahata}, and Y.~{Oyama}.
\newblock {Observation of a neutrino burst from the supernova SN1987A}.
\newblock {\em Phys. Rev. Lett.}, 58:1490, April 1987.

\bibitem{bionta:87}
R.~M. {Bionta}, G.~{Blewitt}, C.~B. {Bratton}, D.~{Casper}, and A.~{Ciocio}.
\newblock {Observation of a neutrino burst in coincidence with supernova 1987A
  in the Large Magellanic Cloud}.
\newblock {\em \prl}, 58:1494, April 1987.

\bibitem{1987ESOC...26..237A}
E.~N. {Alekseev}, L.~N. {Alekseeva}, I.~V. {Krivosheina}, and V.~I.
  {Volchenko}.
\newblock {Detection of the neutrino signal from SN 1987A using the INR Baksan
  underground scintillation telescope}.
\newblock 26:237--247, 1987.

\bibitem{Janka:2012wk}
Hans-Thomas Janka.
\newblock {Explosion Mechanisms of Core-Collapse Supernovae}.
\newblock {\em Ann. Rev. Nucl. Part. Sci.}, 62:407--451, 2012.

\bibitem{Mezzacappa_2024}
Anthony {Mezzacappa} and Michele {Zanolin}.
\newblock {Gravitational Waves from Neutrino-Driven Core Collapse Supernovae:
  Predictions, Detection, and Parameter Estimation}.
\newblock {\em arXiv e-prints}, page arXiv:2401.11635, January 2024.

\bibitem{Szczepanczyk:2021bka}
Marek Szczepanczyk et~al.
\newblock {Detecting and reconstructing gravitational waves from the next
  galactic core-collapse supernova in the advanced detector era}.
\newblock {\em Phys. Rev. D}, 104(10):102002, 2021.

\bibitem{Abbott:2016tdt}
B.P. Abbott et~al.
\newblock {A First Targeted Search for Gravitational-Wave Bursts from
  Core-Collapse Supernovae in Data of First-Generation Laser Interferometer
  Detectors}.
\newblock {\em Phys. Rev. D}, 94(10):102001, 2016.

\bibitem{Abbott:2019pxc}
B.P. Abbott et~al.
\newblock {Optically targeted search for gravitational waves emitted by
  core-collapse supernovae during the first and second observing runs of
  advanced LIGO and advanced Virgo}.
\newblock {\em Phys. Rev. D}, 101(8):084002, 2020.

\bibitem{Szczepanczyk:2023ihe}
Marek~J. Szczepa\'nczyk et~al.
\newblock {Optically targeted search for gravitational waves emitted by
  core-collapse supernovae during the third observing run of Advanced LIGO and
  Advanced Virgo}.
\newblock {\em Phys. Rev. D}, 110(4):042007, 2024.

\bibitem{2023TNSTR1158....1I}
K.~{Itagaki}.
\newblock {Transient Discovery Report for 2023-05-19}.
\newblock {\em Transient Name Server Discovery Report}, 2023-1158:1, May 2023.

\bibitem{cwb}
S.~Klimenko et~al.
\newblock Method for detection and reconstruction of gravitational wave
  transients with networks of advanced detectors.
\newblock {\em Phys. Rev. D}, 93:042004, Feb 2016.

\bibitem{Klimenko:2022nji}
Sergey Klimenko.
\newblock {Wavescan: multiresolution regression of gravitational-wave data}.
\newblock 1 2022.

\bibitem{Shibagaki:20}
Shota {Shibagaki}, Takami {Kuroda}, Kei {Kotake}, and Tomoya {Takiwaki}.
\newblock {A new gravitational-wave signature of low-T/|W| instability in
  rapidly rotating stellar core collapse}.
\newblock {\em Mon. Not. Roy. Astron. Soc.}, 493(1):L138--L142, March 2020.

\bibitem{Bugli:2019rax}
M.~Bugli, J.~Guilet, M.~Obergaulinger, P.~Cerd\'a-Dur\'an, and M.~\'A. Aloy.
\newblock {The impact of non-dipolar magnetic fields in core-collapse
  supernovae}.
\newblock {\em Mon. Not. Roy. Astron. Soc.}, 492(1):58--71, 2020.

\bibitem{Lorimer:2007qn}
D.~R. Lorimer, M.~Bailes, M.~A. McLaughlin, D.~J. Narkevic, and F.~Crawford.
\newblock {A bright millisecond radio burst of extragalactic origin}.
\newblock {\em Science}, 318:777, 2007.

\bibitem{Luck:2010rt}
Harald Luck et~al.
\newblock {The upgrade of GEO600}.
\newblock {\em J. Phys. Conf. Ser.}, 228:012012, 2010.

\bibitem{Affeldt:2014rza}
C.~Affeldt et~al.
\newblock {Advanced techniques in GEO 600}.
\newblock {\em Class. Quant. Grav.}, 31(22):224002, 2014.

\bibitem{Dooley:2015fpa}
K.~L. Dooley et~al.
\newblock {GEO 600 and the GEO-HF upgrade program: successes and challenges}.
\newblock {\em Class. Quant. Grav.}, 33:075009, 2016.

\bibitem{Sutton:2009gi}
Patrick~J. Sutton et~al.
\newblock {X-Pipeline: An Analysis package for autonomous gravitational-wave
  burst searches}.
\newblock {\em New J. Phys.}, 12:053034, 2010.

\bibitem{Was:2012zq}
Michal Was, Patrick~J. Sutton, Gareth Jones, and Isabel Leonor.
\newblock {Performance of an externally triggered gravitational-wave burst
  search}.
\newblock {\em Phys. Rev. D}, 86:022003, 2012.

\bibitem{2021PhRvD.104j2005M}
A.~{Macquet}, M.~A. {Bizouard}, N.~{Christensen}, and M.~{Coughlin}.
\newblock {Long-duration transient gravitational-wave search pipeline}.
\newblock {\em \prd}, 104(10):102005, November 2021.

\bibitem{LIGOScientific:2022jpr}
R.~Abbott et~al.
\newblock {Search for Gravitational Waves Associated with Fast Radio Bursts
  Detected by CHIME/FRB during the LIGO\textendash{}Virgo Observing Run O3a}.
\newblock {\em Astrophys. J.}, 955(2):155, 2023.

\end{thebibliography}
